# Odd-parity multipole order in the spin–orbit-coupled metallic pyrochlore Pb$_2$Re$_2$O$_{7-\delta}$

Yuki Nakayama[1], Daigorou Hirai[1], Hajime Sagayama[2], Keita Kojima[1], Naoyuki Katayama[1],

Jannis Lehmann[3,4], Ziqian Wang[3], Naoki Ogawa[3] and Koshi Takenaka[1]

[1]*Department of Applied Physics, Nagoya University, Nagoya 464–8603, Japan*
[2]*Institute of Materials Structure Science, High Energy Accelerator Research Organization, Tsukuba, Ibaraki 305-0801, Japan*
[3]*RIKEN Center for Emergent Matter Science (CEMS), Wako 351-0198, Japan.*
[4]*Department of Physics, ETH Zurich, 8093 Zurich, Switzerland*

The pyrochlore oxide Pb$_2$Re$_2$O$_{7-\delta}$ (PRO) is a candidate spin–orbit-coupled metal (SOCM) that exhibits a structural phase transition with inversion symmetry breaking. In this study, we report the results of detailed X-ray diffraction (XRD) measurements on single crystals of PRO to clarify the crystal structure below the phase transition temperature at $T_s$ = 300 K. In the XRD patterns, a clear peak splitting is observed below $T_s$, indicating a cubic to tetragonal transition. Based on the group-subgroup relationship and the observed reflection conditions, the space group of the low-temperature phase is proposed to be $I4_122$, which agrees with optical second harmonic generation measurements. This space group is the same as that of the lowest temperature structure of the analogous SOCM Cd$_2$Re$_2$O$_7$ (CRO), which is realized by the emergence of odd-parity multipole order. The comparison between PRO and CRO allows for advancing our understanding on the symmetry-lowering complex order exhibited by SOCMs.

## I. INTRODUCTION

Transition metal compounds containing 4$d$ and 5$d$ electrons have attracted attention for their novel physical properties resulting from their strong spin–orbit interaction (SOI) [1,2]. For example, in Sr$_2$IrO$_4$, the SOI leads to an antiferromagnetic Mott insulating state [3,4]. Due to the similarity of its band structure with that of cuprate superconductors, doping into this SOI-induced Mott insulating state is expected to produce high-temperature superconductivity [5]. The $J$ = 1/2 electronic state with entangled spin and orbital angular momenta has also been exploited for the realization of Kitaev spin liquids [6,7]. Remarkably, the spin–orbit-coupled electronic state gives rise to multipole orders that dominate over the much more common dipolar orders [8–11].

L. Fu pointed out that SOI induces not only Mott insulating state, but also the Fermi-liquid instability in metals, leading to the formation of various electronic phases, and proposed the concept of spin–orbit-coupled metals (SOCM) [12]. A SOCM has a centrosymmetric crystal structure, while the strong SOIs acting on the conduction electrons induces a structural phase transition with spontaneous spatial-inversion-symmetry breaking (ISB). The ISB causes a spin splitting on the Fermi surface (FS), which is expected to promote unconventional odd-parity multipole ordering [13] and exotic superconductivity with parity mixing [14,15]. To date, the candidate compounds of SOCMs are limited to Cd$_2$Re$_2$O$_7$ (CRO) [16], LiOsO$_3$ [17,18], and PbRe$_2$O$_6$ [19]. Although variety of electronic orders such as gyrotropic, ferroelectric and multipolar orders have been theoretically proposed in SOCMs [12], only few of them have been experimentally verified. Progress in material development is key to reveal the diversity of electronic orders formed in SOCMs and further scrutinize the physics of SOCMs.

So far, many studies have focused on CRO, the most investigated candidate for a SOCM [16]. CRO has a cubic α-pyrochlore structure with inversion symmetry at room temperature and has been reported to exhibit metallic conductivity of 5$d$ electrons with strong SOI [20–22].

Three successive structural phase transitions occur in CRO as a function of temperature [23,24]. In the high-temperature phase I, CRO crystallizes in a cubic $Fd\bar{3}m$ structure. First, an ISB transition occurs at $T_{s1}$ = 201.5 K into the tetragonal phase II of space group $I\bar{4}m2$ [24–27]. Second, a transition to an orthorhombic phase XI of space group $F222$ occurs at around $T_{s2}$ ~ 120 K. Third, CRO undergoes a final phase transition to phase III of space group $I4_122$ at $T_{s3}$ ~ 100 K [24].

The magnitude of the tetragonal distortion in CRO below $T_{s1}$ is very small, 0.10% at most [24,28,29]. In contrast to the small structural change, the electronic state clearly changes below $T_{s1}$. The large decrease in the Pauli paramagnetic susceptibility indicates that the density of states decreases by approximately 50% [14]. Hence, this phase transition is not driven by a simple structural instability, but rather by the Fermi liquid instability manifesting in SOCMs [10,11,19]. A theoretical work suggests that, in the phases II and III of CRO, the electronic order associated with ISB can be described by an electric toroidal quadrupole (ETQ) [13]. This unconventional odd-parity multipole ordering induces phenomena such as the spin-split Fermi surface, the magneto-current effect, and nonreciprocal transport in an applied magnetic field [13].

The relevant rhenium-containing pyrochlore oxide, Pb$_2$Re$_2$O$_{7-\delta}$ (PRO), has been reported to exhibit an ISB phase transition from a cubic α-pyrochlore structure (phase I, space group $Fd\bar{3}m$) to a low temperature noncentrosymmetric structure (phase II) at $T_s$ = 300 K [30,31]. The electrical resistivity of PRO indicates metallic behavior in the entire temperature range below 350 K. The small and positive magnetic susceptibility in phase I indicates Pauli paramagnetism, which starts to decrease below $T_s$, similar to the magnetic susceptibility of CRO. The transition at $T_s$ is of second order, as demonstrated by previous heat-capacity measurements [30,31]. The entropy change observed at $T_s$ is $\Delta S$ = 2.4 J K$^{-1}$ mol$^{-1}$ [31], as large as 3.5 J K$^{-1}$ mol$^{-1}$ observed in CRO [32], suggesting that an electronic order is formed as in CRO. Based on the ISB transition and similar physical properties to CRO, PRO likely to be a SOCM However, the formation process of electronic order in CRO, which shows three sequential transitions, and PRO, which shows only one transition, must be significantly different.

For the crystal structure of phase II, K. Ohgushi et al. proposed a cubic space group $F\bar{4}3m$, based on the observed 2$l$ 0 0 superlattice peaks in transmission electron microscopy and single-crystal X-ray diffraction (XRD) measurements [30]. Yet, C. Michioka et al. proposed a tetragonal $I\bar{4}m2$ structure based on the temperature dependence of the 0 0 2 superlattice peak in powder XRD patterns [31], supposing an analogy to the transition of CRO from phase I to II. However, splitting of the diffraction peak, which should be accompanied with the cubic to tetragonal transition, has not been observed. Both studies have reported crystal structures that differ from that of CRO in the lowest temperature phase, which suggest that a different multipole order emerges in PRO.

In this study, we investigated the low temperature phase of PRO, a SOCM candidate. The details of the structural phase transition were investigated by temperature-dependent resistivity, magnetization, and XRD measurements. Electrical-conductivity measurements on single crystals of PRO confirmed metallic conduction between 2 and 350 K. A decrease of the magnetic susceptibility as much as 45% below $T_s$ = 300 K was observed in magnetization measurements. Single-crystal XRD measurements revealed an appearance of superlattice peaks and tetragonal distortion in phase II. Based on the observed reflection conditions and the group-subgroup relationship, we propose space group $I4_122$ for the phase II. Both, the ISB phase transition at $T_s$ = 300 K as well as the derivation of space group $I4_122$ is further supported by optical second harmonic generation (SHG) measurements performed on an as-grown (111) crystal surface. The proposed space group is the same as that of CRO in the lowest temperature phase (phase III), which is described by the $G_u$-type ETQ order.

## II. EXPERIMENTAL

Single crystals of PRO were obtained by conventional solid-state reaction, as reported previously [30]. PbO and ReO$_2$ were mixed in a 1:1 molar ratio and placed in an alumina tube, which was vacuum-sealed in a quartz tube. The tube was heated up to 623 K in 2 h and then to 973 K in 24 h where we kept the temperature for 12 h. We obtained octahedral-shaped single crystals with a purple-metallic luster and a maximum size of 0.3$^3$ mm$^3$.

The magnetic susceptibility was measured with an applied magnetic field ($B$) of 7 T in a magnetic property measurement system (MPMS3; Quantum Design). Electrical resistivity and heat capacity were measured in a physical properties measurement system (PPMS; Quantum Design). Electrical resistivity was measured using the four-probe method. Heat capacity was measured using the relaxation method.

Synchrotron XRD measurements at an energy of 40 keV were performed at the SPring-8 beamline BL02B1. 50 μm-sized single crystals were measured at 350 K and 104 K using an N$_2$ gas blower. The intensities of Bragg reflections were analyzed by *CrysAlisPro* [33]. The intensities of equivalent reflections were averaged by *SORTAV* [34], and the structural parameters were refined by using *Jana2006* [35]. *VESTA* [36] was used to draw the resulting crystal structure.

Synchrotron XRD measurements at Photon Factory, KEK were performed by using an imaging plate (IP) X-ray diffractometer at the beamline BL-8B. An approximately 0.2$^3$ mm$^3$-sized PRO single crystal with a truncated octahedral shape was attached to a glass capillary. The sample temperature was controlled by an N$_2$ gas blower. To increase spatial resolution in the reciprocal space, peak profiles of (20 0 0) and (18 0 0) reflections located in high angle region (diffraction angles larger than 100°) were observed using X-rays with a wavelength of 0.93 Å (13.3 keV).

We performed polarization-resolved SHG measurements in reflection geometry to confirm the ISB and to support our derivation of the space group below $T_s$. We used a femtosecond laser system (LightConversion PHAROS-SP with 6 kHz repetition rate and about 190 fs pulse length) with an optical parametric amplifier (LightConversion ORPHEUS-F). The sample was glued with GE varnish onto a sapphire plate, which was attached with silver paste onto the cold finger of a microscopy cryostat (Janis ST500). We used an objective lens (Olympus SLMPLN50x) to illuminate the sample with our laser beam (wavelength of 800 nm, power of less than 640 μW, spot diameter of approx. 30μm) reflected at 45° incidence from a dichroic beam splitter (Semrock FF700). The incident polarization is adjusted by using a Glan-Laser prism and a half-wave plate. We imaged the sample through the dichroic beam splitter onto a TEC-cooled EM-CCD camera (Hamamatsu C9100-13). Optical filters (two 3-mm-thick Schott BG39 and one 3-mm-thick RG750) and an analyzer (Thorlabs LPVISE100) were used to select SHG at 400 nm wavelength and to define its polarization state, respectively. Each image, from which we finally extract the average SHG intensity, was acquired with 1 minute integration time.

## III. RESULTS

### A. Physical properties

The electrical resistivity and magnetic susceptibility of PRO single crystals as a function of temperature are shown in Fig.1. The electrical resistivity confirms the results of previous studies [30,31] as PRO shows metallic behavior in the entire temperature range below 350 K. The residual resistivity ratio, defined as the ratio of resistivity at 300 and 2 K ($\rho_{300\,K}/\rho_{2\,K}$), is approximately 2, which is in agreement with a previous study [31]. The magnetic susceptibility points towards temperature-independent Pauli paramagnetic behavior with $\chi = 5 \times 10^{-4}$ cm$^3$ mol$^{-1}$ at high temperatures (< 300 K) and gradually decreases to $2.7 \times 10^{-4}$ cm$^3$ mol$^{-1}$ (~100 K). The reduction of magnetic susceptibility starting at around $T_s$ is consistent with previous observations [30,31] and resembles that of CRO. Since the Pauli paramagnetic susceptibility is proportional to the density of states at the Fermi energy, 45% decrease in susceptibility suggests a large change in the electronic structure in the low temperature phase. The observed increase in susceptibility below 50 K is likely attributed to an extrinsic Curie component originating from magnetic impurities (a few percent, assuming $S = 1/2$).

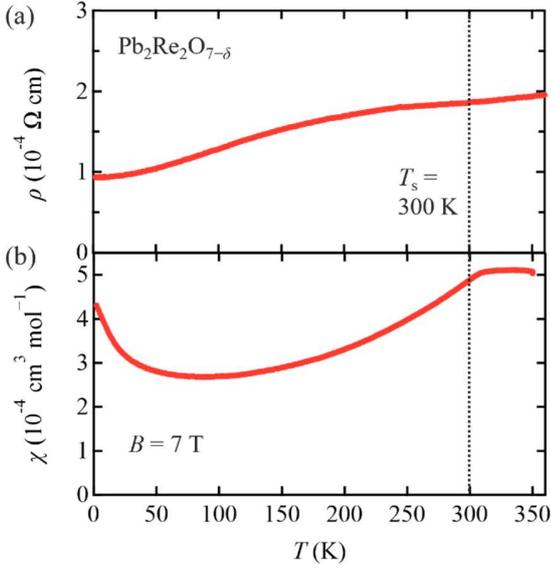

FIG. 1. Temperature dependence of (a) the electrical resistivity $\rho$ and (b) the magnetic susceptibility $\chi$ of Pb$_2$Re$_2$O$_{7-\delta}$ single crystals. The phase transition temperature $T_s$ = 300 K is indicated as a vertical dotted line.

### B. Single crystal X-ray diffraction

To clarify the structural changes at the phase transition $T_s$, synchrotron single-crystal-XRD experiments were performed. The unwarped diffraction data of phase I at 350 K (measured at BL02B1 of SPring-8) is shown in Fig. 2(a). The diffraction data show the presence of the $h00$: $h = 2n+1$ extinction rule originating from the face-centered lattice and the $h00$: $h = 4n+2$ extinction rule originating from the $d$-glide plane. Therefore, our refinement assumes an α-pyrochlore structure with the space group $Fd\bar{3}m$, confirming previous reports [30]. In the α-pyrochlore structure, Pb and Re occupy one site each, whereas O occupies two different sites. One O site forms the ReO$_6$ octahedron, and the other O site (O') is located at the center of the Pb$_4$ tetrahedron. We performed a structural analysis based on the composition ratio of Pb$_2$Re$_2$O$_{6.7}$ obtained by iodine titration and compositional analysis as reported previously [30]. As a result of the assignment of the deficiency to the O and O' sites, the best $R$ value ($R_{all}$=1.38%) was obtained assuming Pb$_2$Re$_2$O$_6$O'$_{0.7}$, in agreement with the previous study. The corresponding CIF file is added as a Supplemental Material [37].

The unwarped diffraction data of phase II, shown in Fig. 2(b), confirms the violation of the $d$-glide plane derived extinction rule, which was present in phase I. This indicates that a structural phase transition has occurred, as concluded in previous studies [30,31]. The observation of anomalous anisotropic temperature factors for the oxygen sites in the refined structure (Fig. S1 in Supplemental Material [37]) rules out the two cubic space groups ($F\bar{4}3m$ and $F23$) in the $Fd\bar{3}m$ subgroup that also satisfy the extinction rule. Thus, the crystal system in Phase II is likely non-cubic. The diffraction image in Fig. 2(b) does not clearly show the splitting of the Bragg peak associated with a symmetry lowering, indicating that the lattice parameters of phase II deviate only slightly from those of a cubic lattice.

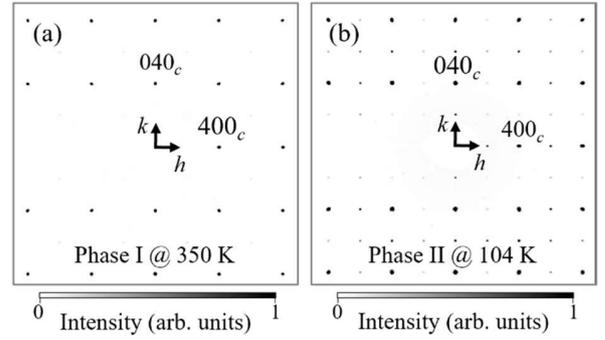

FIG. 2. Unwarped diffraction data of the $hk0$ plane for Pb$_2$Re$_2$O$_{7-\delta}$ (a) in phase I, at 350 K and (b) in phase II, at 104 K. The appearance of superlattice peaks in (b) indicates the violation of the $d$-glide plane derived extinction rule.

### C. Structural change

To obtain more evidence for a lower crystal symmetry, we performed high-resolution single-crystal XRD at large diffraction angles up to 140° at the beamline BL-8B (KEK Photon Factory). We probed the temperature dependence of the two Bragg reflections (18 0 0) and (20 0 0) in phase I (cubic α-pyrochlore), as shown in Fig. 3. Below $T_s$, the (18 0 0) reflection emerges upon cooling, which indicates the violation of the $h00$: $h = 4n+2$ extinction rule originating from the $d$-glide plane, as is consistent with the XRD results from SPring-8. In addition, the (20 0 0) reflection splits into two peaks: a low-angle and a high-angle peaks at an intensity ratio of approximately 2:1. It is noted that the (18 0 0) reflection is a single peak, not a double peak like the (20 0 0) reflection, demonstrating the presence of an additional extinction rule in phase II.

When a structural phase transition from cubic to tetragonal occurs, three types of domains are generally formed with the tetragonal $c$-axis parallel to the three directions [1 0 0], [0 1 0] and [0 0 1]. The (20 0 0) and (0 20 0) reflections from these three types of domains are observed at the same diffraction angle, but only the reflection corresponding to (0 0 20) appears at a different angle due to $a \neq c$. Assuming equal volume fractionations of each domain, all the reflections should have the same intensity. Thus, the ratio of the two peak [(20 0 0) + (0 20 0) and (0 0 20)] intensities should be 2:1. At 120 K the ratio is 1.8:1, which is reasonably close to the theoritical value.

As phase II is suggested to form a bct lattice [38], the two peaks emerging from the $(20\ 0\ 0)_c$ reflection are indexed as $(10\ 10\ 0)_t$ and $(0\ 0\ 20)_t$ reflections for the low- and high-angle peaks, respectively. Therefore, we conclude that $\sqrt{2}a_t > c_t$ in the tetragonal phase II. Furthermore the superlattice reflection appearing at around the $(18\ 0\ 0)_c$ reflection is identified as a $(9\ 9\ 0)_t$ reflection using the lattice constants obtained from the aforementioned $(10\ 10\ 0)_t$ and $(0\ 0\ 20)_t$ reflections. The appearance of the $(9\ 9\ 0)_t$ reflection indicates the violation of the $hkl$: $h + k + l = 4n+2$ extinction rule derived from the $d$-glide plane. The absence of the $(0\ 0\ 18)_t$ reflection near the $(9\ 9\ 0)_t$ reflection indicates the additional extinction rule $00l$ : $l = 4n + 2$, likely originating from a $4_1$ ($4_3$) screw axis parallel to the $c$ direction.

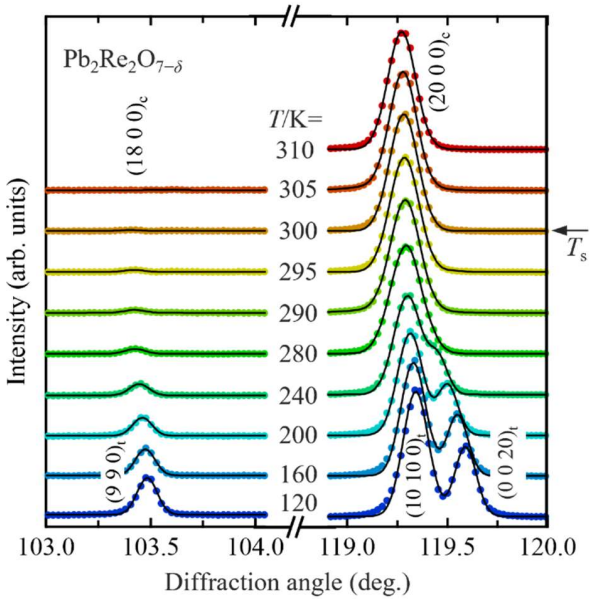

FIG. 3. Temperature variation of XRD patterns near the $(20\ 0\ 0)_c$ and $(18\ 0\ 0)_c$ reflections of PRO. Solid black lines represent single Gaussian fits above 305 K and double-Gaussian fits below 300 K for the $(20\ 0\ 0)_c$ reflection and as well as only single Gaussian fits for the $(18\ 0\ 0)_c$ reflection. The transition temperature $T_s$ = 300 K is marked with an arrow.

The temperature dependence of the full width at half maximum (FWHM) of the $(20\ 0\ 0)_c$ reflection, obtained by fitting with a single Gaussian, is shown in Fig. 4(a). The FWHM is nearly constant above 300 K, and increases with decreasing temperature below 300 K. This is attributed to the onset of a peak splitting due to the cubic-to-tetragonal transition occurring at $T_s$ = 300 K. The deviation of the lattice parameter from the cubic unit cell ($\sqrt{2}a_t - c_t$) is also shown in Fig. 4(a). The temperature dependence of the lattice constants was determined using the peak center obtained from fittings with a single Gaussian above 305 K and by a double Gaussian below 300 K. The smooth development of $\sqrt{2}a_t - c_t$ suggests that the phase transition at $T_s$ is of second order, as supported by previous heat-capacity measurements [30,31] and by the temperature dependence of the (0 0 2) reflection intensities measured on the powder samples [31].

The temperature dependence of the $(9\ 9\ 0)_t$ reflection intensity is shown in Fig. 4(b). By fitting the temperature-dependent intensity from 235 to 300 K to the power function $(T_s - T)^\beta$, the transition temperature $T_s$ = 302(4) K and the critical exponent $\beta$ = 0.85(15) are obtained. The intensity of the superlattice reflection starts increasing at approximately the same temperature ($T$ = 300 K) as the increase of the FWHM of the $(20\ 0\ 0)_c$ reflection. Therefore, the tetragonal distortion and the loss of $d$-glide plane symmetry occur simultaneously at $T_s$, within the experimental error.

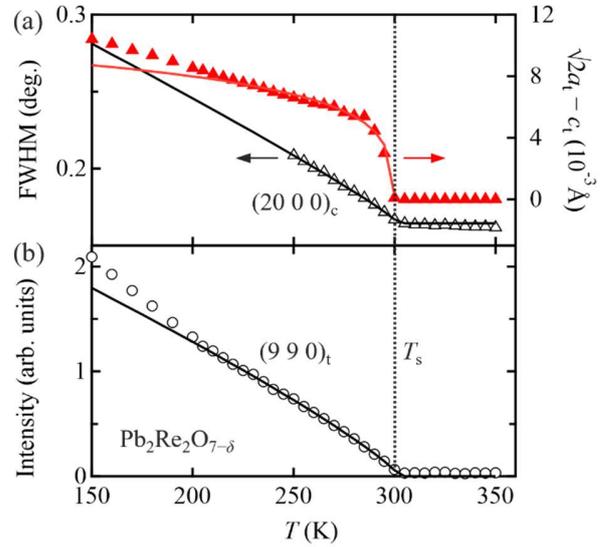

FIG. 4. (a) Temperature dependence of the FWHM of the $(20\ 0\ 0)_c$ reflection and the tetragonal distortion ($\sqrt{2}a_t - c_t$) of PRO, and (b) the intensity of $(9\ 9\ 0)_t$ reflection. The solid lines are the fit following the temperature dependence as $(T_s - T)^\beta$. $T_s$ is shown as a vertical dotted line.

In Fig.5 we show the temperature dependence of the lattice constants and unit-cell volume obtained by the center of Gaussians for the $(20\ 0\ 0)_c$, $(10\ 10\ 0)_t$, and $(0\ 0\ 20)_t$ reflections. In both phases, the lattice constants decrease with decreasing temperature. As a result, the unit cell volume also decreases monotonically upon cooling in the measured temperature range between 350 and 120 K. This behavior contrasts with the one of CRO, where the unit-cell volume increases with decreasing temperature below the ISB phase transition at $T_{s1}$ [24].

The tetragonal distortion ($\sqrt{2}a_t - c_t$) of PRO is 0.011 Å at 120 K, which is approximately twice that of CRO at 10 K (0.0058 Å) [24]. In contrast, other pyrochlore compounds exhibit greater distortion than that of PRO: $Bi_2Hf_2O_7$, which changes from a cubic α-pyrochlore structure to a tetragonal $I4_122$ structure at 1173 K, exhibits a tetragonal distortion of 0.023 Å at 873 K [39]. The deviation from a cubic lattice (($\sqrt{2}a - b$)/$b$) for $Cd_2Nb_2O_7$ with an orthorhombic $Ima2$ structure is 0.017 Å at 100 K [40]. The tetragonal distortion that accompanies the phase transition in PRO is relatively small despite the large change in electronic state, which supports our assumption that the ISB phase transition is driven by an electronic instability.

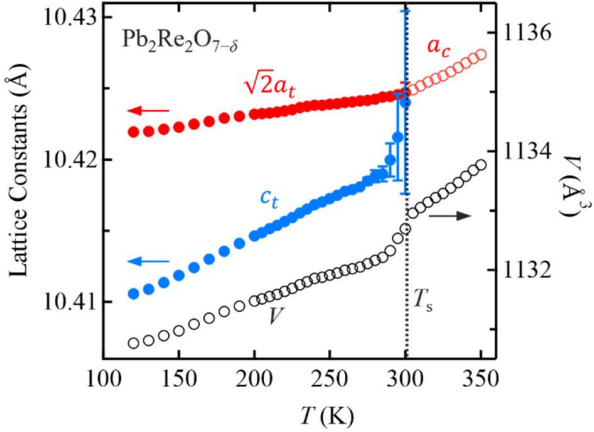

FIG. 5. Temperature dependence of the lattice constants and the unit-cell volume of PRO. We assume a body-centered tetragonal cell in phase II below $T_s$ (vertical dotted line). The errors of the lattice constants become larger when approaching $T_s$ because of the reduced peak splitting.

### D. Low temperature structure

Since the phase transition at $T_s$ is of second order, the group-subgroup relation of the space group is applied to deduce the space group of phase II. In Fig. 6, we illustrates the maximum group-subgroup relations in the fcc and bct lattices starting from $Fd\bar{3}m$. These space groups are compared in Tabel. 1.

First, cubic $F\bar{4}3m$ and $F4_132$ space groups are excluded from candidates because phase II has a tetragonal structure as indicated by the splitting of the (20 0 0)$_c$ reflection. Second, tetragonal space groups without $4_1(4_3)$ screw axis, namely $I\bar{4}2d$, $I\bar{4}m2$, and $I\bar{4}$ are ruled out. The existence of a $4_1(4_3)$ screw axis was confirmed by the absence of 00$l$: $l = 4n + 2$ reflections such as the (0 0 18)$_t$ reflection. Third, space groups with additional glide planes ($I4_1md$, $I4_1/amd$, and $I4_1/a$.) are precluded. For space group $I4_1md$, which has a $d$-glide plane that is parallel to the $4_1$ screw axis, $hkl$: $h + k + l \neq 4n$ reflections are forbidden. The observed (9 9 0)$_t$ reflection is in contradiction with the extinction rule of space group $I4_1md$. For the space groups $I4_1/amd$ and $I4_1/a$ with an $a$-glide plane perpendicular to the $4_1(4_3)$ screw axis, $hk0$: $h = 2n + 1$ reflections must be absent. Again, the presence of the (9 9 0)$_t$ reflection excludes the space groups $I4_1/amd$ and $I4_1/a$.

Accordingly, only two candidate space groups, $I4_122$ and $I4_1$, remain. Because of the higher symmetry with less constrains that are supported by our experimental data, the tetragonal $I4_122$ space group is our proposal for phase II of PRO. This space group is identical to that of the related SOCM, CRO, in the lowest temperature phase (phase III).

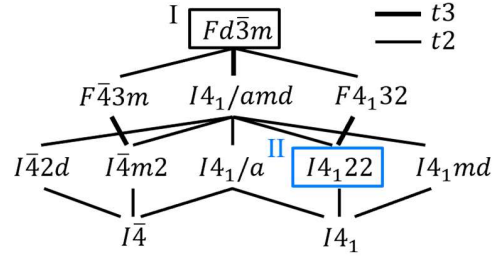

FIG. 6. Maximum group-subgroup relations on face-centered cubic and body-centered tetragonal lattices, starting from $Fd\bar{3}m$. The symmetry reduction from face-centered to body-centered lattice is described by the translational-equivalent index 3($t3$) due to cell change, and the other reduction is described by $t2$.

Tabel 1. Comparison of subgroups on face-centered cubic and body-centered tetragonal lattices with $Fd\bar{3}m$ as the parent space group.

| Space group (#) | Crystal system | $4_1$ screw axis | Reason for being ruled out |
|---|---|---|---|
| $F\bar{4}3m$ (216) | cubic | No | cubic system |
| $F4_132$ (210) | cubic | Yes | cubic system |
| $I\bar{4}2d$ (122) | tetragonal | No | absence of $4_1$ screw axis |
| $I\bar{4}m2$ (119) | tetragonal | No | absence of $4_1$ screw axis |
| $I\bar{4}$ (82) | tetragonal | No | absence of $4_1$ screw axis |
| $I4_1md$ (109) | tetragonal | Yes | $d$-glide plane |
| $I4_1/amd$ (141) | tetragonal | Yes | $a$-glide plane |
| $I4_1/a$ (88) | tetragonal | Yes | $a$-glide plane |
| $I4_122$ (98) | tetragonal | Yes | |
| $I4_1$ (80) | tetragonal | Yes | |

### E. Order parameter

The structural phase transition from space group $Fd\bar{3}m$ to $I4_122$ is induced by a two-dimensional $E_u$ mode as the order parameter (OP) [13]. Another structure induced by the $E_u$ mode is space group $I\bar{4}m2$, which is the structure of phase II of CRO between 201.5 and 115.4 K. The OP of the $E_u$ mode is represented by a two-dimensional vector $\boldsymbol{\eta} = (\eta_1, \eta_2)$, where only $\eta_1$ ($\eta_2$) is finite for space group $I4_122$ ($I\bar{4}m2$). The two-dimensional vector can be expressed in polar coordinates as $\boldsymbol{\eta} = (\eta, \theta)$, $\eta_1 = \eta\cos\theta$, and $\eta_2 = \eta\sin\theta$, where $\eta$ and $\theta$ are the amplitude and phase of the OP, respectively. Considering $\eta$ up to the second order of the free energy, the lattice constants $a_o$, $b_o$, $c_o$ in each direction of the orthorhombic unit cell are as follows.

$$a_o = a_0 + a_A\eta^2 + a_E\eta^2\cos(2\theta + 2\pi/3),$$
$$b_o = a_0 + a_A\eta^2 + a_E\eta^2\cos(2\theta - 2\pi/3),$$
$$c_o = a_0 + a_A\eta^2 + a_E\eta^2\cos(2\theta), \quad (1)$$

, where $a_0$ is a temperature-dependent constant due to thermal expansion and $a_A$ and $a_E$ are temperature-independent constants [24]. Transforming equation (1), we obtain,

$$\tan(2\theta) = \sqrt{3}(b_o - a_o)/(2c_o - a_o - b_o),$$
$$\eta = [(2c_o - a_o - b_o)/3a_E\cos(2\theta)]^{1/2} \quad (2).$$

Substituting the experimentally obtained lattice constants into equation (2), we obtain $2\theta = 0$ for both phase I and phase II. Hence, only $\eta_1$ has a finite value in phase II of space group $I4_122$. The equation (2) requires that $\sqrt{2}a_t > c_t$ for space group $I4_122$ which we have experimentally confirmed.

Shown in Fig. 7 is the temperature dependence of the amplitude of the OP $\eta$, calculated from Eq. (2) with $a_E = 1$. $\eta$ increases sharply at $T_s$, implying that it is the primary OP for the phase transition. Fitting with a power function below 300 K, the transition temperature, yields a critical exponent $\beta = 0.179(14)$. This value is close to $\beta = 0.223(5)$ at $T_{s1}$ of CRO [24] and may indicate that the structural phase transition is promoted by a similar mechanism.

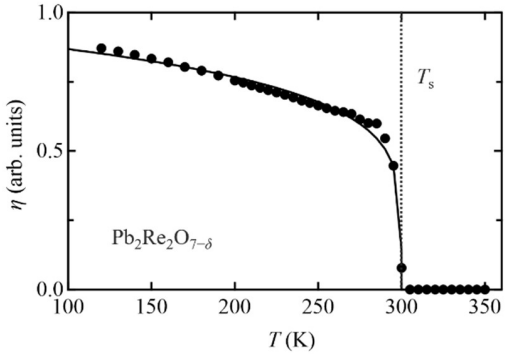

FIG. 7. Temperature dependence of the order-parameter amplitude $\eta$. The solid line results from a fitting with a power function. $T_s$ is shown as a vertical dashed line.

### F. Inversion symmetry breaking

We performed optical SHG measurements to elucidate the temperature dependence and the nature of the ISB in PRO. In the dipole approximation, the generation of a second-harmonic wave is only allowed in crystals without a center of inversion, while moreover, the polarization dependence of the SHG intensity can help to reveal details of the crystals' point-group symmetry [41].

Here, the generation of a nonlinear polarization $P_{2\omega}$ in response to the electric field of an incident light wave $E_\omega$ oscillating at frequency $\omega$, described as $P_{i(2\omega)} \propto \chi_{ijk} E_{j(\omega)} E_{k(\omega)}$, reflects the symmetry properties of the material via its second-order nonlinear-optical susceptibility $\chi_{ijk}$. The components of that $\chi_{ijk}$ tensor couple linearly to the proposed polar order parameter in pyrochlore-type SOCMs. As a consequence, the SHG intensity, which is proportional to the square of the nonlinear polarization, is also proportional to the square of the order parameter. The temperature dependence of the SHG-intensity-deduced order parameter is shown in Fig. 8(a). The increase in SHG intensity below $T_s$ confirms ISB at the transition.

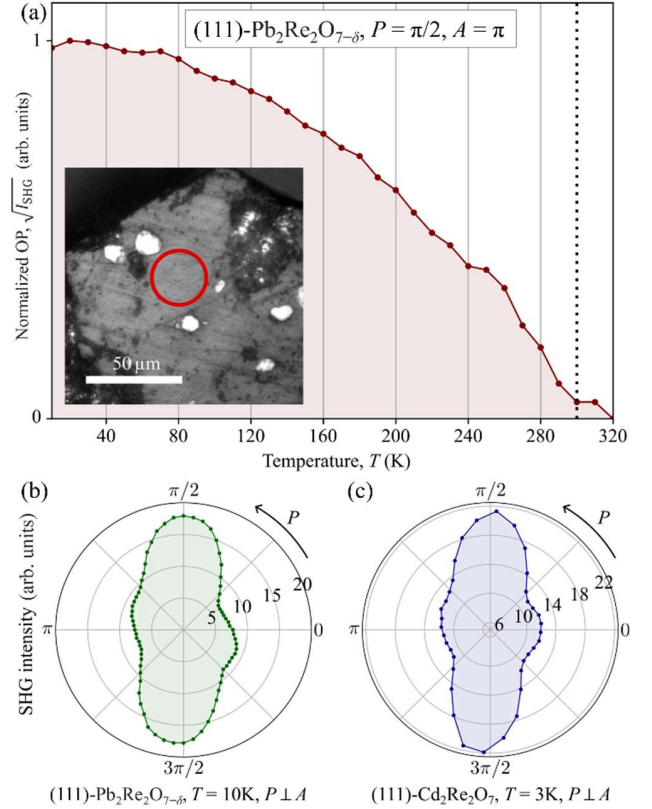

Fig. 8. Temperature- and polarization-dependent optical SHG measurements. $P$ and $A$ denote the angle of polarizations for incident and reflected light, respectively. (a) Order-parameter amplitude derived from the SHG intensity measured on the (111) surface of a PRO single crystal. Shown in the inset is a white-light image of the crystal surface. The bright regions correspond to impurities on the crystal surface. The red circle indicates the exposed area selected for data acquisition. The laser-spot diameter (FWHM) at the sample surface is estimated to be 30 μm. The onset of SHG at around 300 K is due to ISB at the phase transition (vertical dotted line). (b) Rotational-anisotropy SHG measurement with crossed polarizer and analyzer performed on (111)-PRO at $T = 10$ K. The pattern resembles the one measured in the low temperature phase of (111)-CRO at $T = 3$ K (c).

The $\chi$ tensor for the proposed space group $I4_122$ contains just one ($\chi_{xyz} = \chi_{yxz}$) component, while the coexistence of three domain species each with a binary degree of freedom reflecting the polar nature of the suggested order parameter, hinders an unambiguous theoretical modeling of the data, given the lack of information on the domain configuration of our sample. Hence, instead of a theoretical modeling, we compare our SHG results on PRO with data that we acquired on a (111)-CRO surface. Since both compounds are proposed to share the $I4_122$ space group in their low-temperature phase, a qualitatively similar SHG rotational-anisotropy pattern is expected and indeed experimentally observed, see Figs. 8(b) and 8(c).

## IV. DISCUSSION

The proposed tetragonal $I4_122$ structure is shared by the two compounds PRO and CRO in their lowest-temperature phases, which may suggest the same driving force for the phase transitions. In contrast, only a single structural phase transition occurs in PRO, whereas CRO exhibits three successive structural phase transitions [23,24]. The complex phase transitions in CRO may reflect the proximity of degenerate ground states originating from the doubly degenerate $E_u$ OP.

To date, seven pyrochlore oxides containing Pb at the $A$ site have been reported, among which no structural phase transition has been reported except for PRO. Seven of them, $B$ = Nb, Tc, Os, and Ta compounds have α-pyrochlore structures [42–45], while $B$ = Ru and Ir compounds crystallize in the noncentrosymmetric structure of space group $F\bar{4}3m$ [46,47]. Except for PRO and CRO, the only reported pyrochlore oxide crystallizing in a $I4_122$ structure is the γ-$Bi_2Hf_2O_7$ at which the structure is formed between 873 and 1173 K [39]. The origin of the structural transitions of $Bi_2Hf_2O_7$ has been argued to be a stereochemically active lone pair of Bi ions, considering the similarity with another Bi containing pyrochlore oxide $Bi_2Sn_2O_7$ [48,49]; the structural phase transition in PRO is also discussed to be driven by a lone pair of the Pb ion [30]. However, the facts that most of Pb-based pyrochlore oxides do not exhibit phase transitions and that CRO, which contain Cd without lone pairs, and PRO have similar phase transitions suggest that Pb lone pairs do not have a strong influence on the phase transition of PRO.

Another origin of the structural instability in the pyrochlore oxide $A_2Re_2O_6O'$ has been discussed in terms of the covalent nature of the $A$-O' bond [30,40]. Since both Cd and Pb form covalent bonds with oxygen, this supports the common phase transition in CRO and PRO. On the other hand, the fact that several Pb containing pyrochlore compounds have undistorted α-pyrochlore structures also makes it difficult to discuss the stabilization of noncentrosymmetric structures solely by A-O' covalent bonds.

Metals with noncentrosymmetric structures such as the low-temperature phases of PRO and CRO are rare. This is partly because the itinerant electrons screen the crystal-electric fields and inhibit long-range Coulomb interactions responsible for ISB in insulators. Therefore, it is likely that the phase transitions in CRO and PRO are caused by different mechanisms than in insulating pyrochlores such as $Bi_2Hf_2O_7$ [39] and $Cd_2Nb_2O_7$ [50,51] which undergo ISB phase transitions as well. Considering that the density of states decreases significantly accompanied by the phase transition, it is inferred that in the regular structure above the transition temperatures, the electronic state is unstable due to a high density of states. It would be natural to assume that the phase transition is driven by electronic instability.

The electronic order emerging in CRO is considered to be associated with unconventional multipoles, electronic toroidal quadrupole (ETQ) orders [11,19,33–35]. Each electric toroidal moment is described by a circular (toroidal) arrangement of alternating electric charges produced by specific Re-Re bond modulation in the $Re_4$ tetrahedral unit of the pyrochlore structure. Different arrangements of toroidal moments are proposed for phase II and III of CRO: ETQ order with $x^2$-$y^2$ component ($G_v$-type) and $3z^2$-$r^2$ component ($G_u$-type) for phase II and III, respectively.

Since the lowest temperature structure of CRO and PRO is supposed to be identical, $G_u$-type ETQ is likely to be formed in the phase II of PRO. Here, $G_u$-type order is selected as the ground state from the doubly degenerate $G_v$ and $G_u$-type ETQ orders. Electronic details about the origin of this symmetry breaking are of interest for future studies. Further comparison of the ISBs with respect to their origin and final manifestation in CRO and PRO would be useful for fundamental understanding of SOCMs.

## V. CONCLUSIONS

In this study, we characterized $Pb_2Re_2O_{7-\delta}$, a candidate material for SOCM. We uncovered details on the structural phase transition at $T_s$ = 300 K in single crystals by applying electrical resistivity, magnetic susceptibility, detailed XRD, and optical SHG measurements. Single-crystal XRD pattern revealed peak splittings indicating tetragonal distortion and the extinction conditions in the low-temperature phase. Based on our structural analysis, the space group of the low-temperature phase is proposed to be $I4_122$. This space group is also formed in the lowest temperature phase of $Cd_2Re_2O_7$, which is associated with $G_u$-type ETQ order. Due to the similarity of the phase transitions in $Pb_2Re_2O_{7-\delta}$, and $Cd_2Re_2O_7$, suppression of the ISB by applying pressure and emergence of exotic superconductivity may be expected, as in $Cd_2Re_2O_7$ [52]. We propose future experiments targeting the robustness of the phase transition by investigating the impact of oxygen vacancies and by increasing the covalency of the bonds by substituting O with S or Se. Furthermore, the successful growth of crystals with high-quality mm-sized natural facets may open the door for studying the domain configuration via SHG microscopy.

## ACKNOWLEDGMENT

This study was partly conducted under the Visiting Researcher's Program of the Institute for Solid State Physics, the University of Tokyo. The single crystal XRD experiment was conducted at the BL02B1 of SPring-8, Hyogo, Japan (Proposals No. 2022B0607). The high-resolution XRD experiment was performed under the approval of the Photon Factory Program Advisory Committee (Proposal No. 2022G556). Z.W. was supported by RIKEN Incentive Research Projects. This study was partly supported by the Japan Society for the Promotion of Science (JSPS) KAKENHI

Grant No. JP20H01858, JP21K13889, JP22KJ1521, JP22H04462 (Quantum Liquid Crystals), and JP23H04860.

[1] W. Witczak-Krempa, G. Chen, Y. B. Kim, and L. Balents, *Correlated Quantum Phenomena in the Strong Spin-Orbit Regime*, Annu. Rev. Condens. Matter Phys. **5**, 57 (2014).

[2] T. Takayama, J. Chaloupka, A. Smerald, G. Khaliullin, and H. Takagi, *Spin–Orbit-Entangled Electronic Phases in 4d and 5d Transition-Metal Compounds*, J. Phys. Soc. Japan **90**, 062001 (2021).

[3] B. J. Kim, H. Jin, S. J. Moon, J.-Y. Y. Kim, B.-G. G. Park, C. S. Leem, J. Yu, T. W. Noh, C. Kim, S.-J. J. Oh, J.-H. H. Park, V. Durairaj, G. Cao, and E. Rotenberg, *Novel $j_{Eff} = 1/2$ Mott State Induced by Relativistic Spin-Orbit Coupling in $Sr_2IrO_4$*, Phys. Rev. Lett. **101**, 076402 (2008).

[4] B. J. Kim, H. Ohsumi, T. Komesu, S. Sakai, T. Morita, H. Takagi, and T. Arima, *Phase-Sensitive Observation of a Spin-Orbital Mott State in $Sr_2IrO_4$*, Science **323**, 1329 (2009).

[5] F. A. Wang and T. Senthil, *Twisted Hubbard Model for $Sr_2IrO_4$: Magnetism and Possible High Temperature Superconductivity*, Phys. Rev. Lett. **106**, 136402 (2011).

[6] G. Jackeli and G. Khaliullin, *Mott Insulators in the Strong Spin-Orbit Coupling Limit: From Heisenberg to a Quantum Compass and Kitaev Models*, Phys. Rev. Lett. **102**, 017205 (2009).

[7] J. G. Rau, E. K.-H. Lee, and H.-Y. Kee, *Spin-Orbit Physics Giving Rise to Novel Phases in Correlated Systems: Iridates and Related Materials*, Annu. Rev. Condens. Matter Phys. **7**, 195 (2016).

[8] G. Chen, R. Pereira, and L. Balents, *Exotic Phases Induced by Strong Spin-Orbit Coupling in Ordered Double Perovskites*, Phys. Rev. B **82**, 174440 (2010).

[9] D. Hirai and Z. Hiroi, *Successive Symmetry Breaking in a $J_{Eff} = 3/2$ Quartet in the Spin–Orbit Coupled Insulator $Ba_2MgReO_6$*, J. Phys. Soc. Japan **88**, 064712 (2019).

[10] D. Hirai, H. Sagayama, S. Gao, H. Ohsumi, G. Chen, T. Arima, and Z. Hiroi, *Detection of Multipolar Orders in the Spin-Orbit-Coupled 5d Mott Insulator $Ba_2MgReO_6$*, Phys. Rev. Res. **2**, 022063 (2020).

[11] D. D. Maharaj, G. Sala, M. B. Stone, E. Kermarrec, C. Ritter, F. Fauth, C. A. Marjerrison, J. E. Greedan, A. Paramekanti, and B. D. Gaulin, *Octupolar versus Néel Order in Cubic $5d^2$ Double Perovskites*, Phys. Rev. Lett. **124**, 087206 (2020).

[12] L. Fu, *Parity-Breaking Phases of Spin-Orbit-Coupled Metals with Gyrotropic, Ferroelectric, and Multipolar Orders*, Phys. Rev. Lett. **115**, 026401 (2015).

[13] S. Hayami, Y. Yanagi, H. Kusunose, and Y. Motome, *Electric Toroidal Quadrupoles in the Spin-Orbit-Coupled Metal $Cd_2Re_2O_7$*, Phys. Rev. Lett. **122**, 147602 (2019).

[14] V. Kozii and L. Fu, *Odd-Parity Superconductivity in the Vicinity of Inversion Symmetry Breaking in Spin-Orbit-Coupled Systems*, Phys. Rev. Lett. **115**, 207002 (2015).

[15] Y. Wang, G. Y. Cho, T. L. Hughes, and E. Fradkin, *Topological Superconducting Phases from Inversion Symmetry Breaking Order in Spin-Orbit-Coupled Systems*, Phys. Rev. B **93**, 134512 (2016).

[16] Z. Hiroi, J.-I. Yamaura, T. C. Kobayashi, Y. Matsubayashi, and D. Hirai, *Pyrochlore Oxide Superconductor $Cd_2Re_2O_7$ Revisited*, J. Phys. Soc. Japan **87**, 024702 (2018).

[17] Y. Shi, Y. Guo, X. Wang, A. J. Princep, D. Khalyavin, P. Manuel, Y. Michiue, A. Sato, K. Tsuda, S. Yu, M. Arai, Y. Shirako, M. Akaogi, N. Wang, K. Yamaura, and A. T. Boothroyd, *A Ferroelectric-like Structural Transition in a Metal*, Nat. Mater. **12**, 1024 (2013).

[18] G. Giovannetti and M. Capone, *Dual Nature of the Ferroelectric and Metallic State in $LiOsO_3$*, Phys. Rev. B **90**, 195113 (2014).

[19] S. Tajima, D. Hirai, T. Yajima, D. Nishio-Hamane, Y. Matsubayashi, and Z. Hiroi, *Spin–Orbit-Coupled Metal Candidate $PbRe_2O_6$*, J. Solid State Chem. **288**, 121359 (2020).

[20] D. J. Singh, P. Blaha, K. Schwarz, and J. O. Sofo, *Electronic Structure of the Pyrochlore Metals $Cd_2Os_2O_7$ and $Cd_2Re_2O_7$*, Phys. Rev. B **65**, 155109 (2002).

[21] Y. Matsubayashi, K. Sugii, H. T. Hirose, D. Hirai, S. Sugiura, T. Terashima, S. Uji, and Z. Hiroi, *Split Fermi Surfaces of the Spin–Orbit-Coupled Metal $Cd_2Re_2O_7$ Probed by de Haas–van Alphen Effect*, J. Phys. Soc. Japan **87**, 053702 (2018).

[22] H. T. Hirose, T. Terashima, D. Hirai, Y. Matsubayashi, N. Kikugawa, D. Graf, K. Sugii, S. Sugiura, Z. Hiroi, and S. Uji, *Electronic States of Metallic Electric Toroidal Quadrupole Order in $Cd_2Re_2O_7$ Determined by Combining Quantum*

Oscillations and Electronic Structure Calculations, Phys. Rev. B **105**, 035116 (2022).

[23] S. Uji, H. T. Hirose, T. Terashima, Y. Matsubayashi, D. Hirai, Z. Hiroi, and T. Hasegawa, *Successive Continuous Phase Transitions in Spin–Orbit Coupled Metal $Cd_2Re_2O_7$*, J. Phys. Soc. Japan **90**, 064714 (2021).

[24] D. Hirai, A. Fukui, H. Sagayama, T. Hasegawa, and Z. Hiroi, *Successive Phase Transitions of the Spin-Orbit-Coupled Metal $Cd_2Re_2O_7$ Probed by High-Resolution Synchrotron x-Ray Diffraction*, J. Phys. Condens. Matter **35**, 35403 (2023).

[25] C. A. Kendziora, I. A. Sergienko, R. Jin, J. He, V. Keppens, B. C. Sales, and D. Mandrus, *Goldstone-Mode Phonon Dynamics in the Pyrochlore $Cd_2Re_2O_7$*, Phys. Rev. Lett. **95**, 125503 (2005).

[26] J. C. Petersen, M. D. Caswell, J. S. Dodge, I. A. Sergienko, J. He, R. Jin, and D. Mandrus, *Nonlinear Optical Signatures of the Tensor Order in $Cd_2Re_2O_7$*, Nat. Phys. **2**, 605 (2006).

[27] J. W. Harter, Z. Y. Zhao, J. Q. Yan, D. G. Mandrus, and D. Hsieh, *A Parity-Breaking Electronic Nematic Phase Transition in the Spin-Orbit Coupled Metal $Cd_2Re_2O_7$*, Science **356**, 295 (2017).

[28] J. P. Castellan, B. D. Gaulin, J. van Duijn, M. J. Lewis, M. D. Lumsden, R. Jin, J. He, S. E. Nagler, and D. Mandrus, *Structural Ordering and Symmetry Breaking in $Cd_2Re_2O_7$*, Phys. Rev. B **66**, 134528 (2002).

[29] J.-I. Yamaura and Z. Hiroi, *Low Temperature Symmetry of Pyrochlore Oxide $Cd_2Re_2O_7$*, J. Phys. Soc. Japan **71**, 2598 (2002).

[30] K. Ohgushi, J. I. Yamaura, M. Ichihara, Y. Kiuchi, T. Tayama, T. Sakakibara, H. Gotou, T. Yagi, and Y. Ueda, *Structural and Electronic Properties of Pyrochlore-Type $A_2Re_2O_7$ (A = Ca, Cd, and Pb)*, Phys. Rev. B **83**, 125103 (2011).

[31] C. Michioka, Y. Kataoka, H. Ohta, and K. Yoshimura, *Possible Spin Singlet Quadrumerization in $Pb_2Re_2O_{7-\delta}$*, J. Phys. Condens. Matter **23**, 445602 (2011).

[32] Z. Hiroi, J. I. Yamaura, Y. Muraoka, and M. Hanawa, *Second Phase Transition in Pyrochlore Oxide $Cd_2Re_2O_7$*, J. Phys. Soc. Japan **71**, 1634 (2002).

[33] C. P. R. O. Agilent and P. R. O. CrysAlis, *Agilent Technologies Ltd*, Yarnton, Oxfordshire, Engl. (2014).

[34] R. H. Blessing, *Data Reduction and Error Analysis for Accurate Single Crystal Diffraction Intensities*, Crystallogr. Rev. **1**, 3 (1987).

[35] V. Petříček, M. Dušek, and L. Palatinus, *Crystallographic Computing System JANA2006: General Features*, **229**, 345 (2014).

[36] K. Momma and F. Izumi, *VESTA 3 for Three-Dimensional Visualization of Crystal, Volumetric and Morphology Data*, J. Appl. Crystallogr. **44**, 1272 (2011).

[37] See Supplemental Material below for the crystal information file of high temperature phase of $Pb_2Re_2O_{7-\delta}$ and structural analysis for phase II, assuming a cubic space group.

[38] The primitive lattice vectors of a body-centered tetragonal (bct) lattice emerging from a face-centered cubic (fcc) lattice are represented by $a_t = (a_c - b_c)/2$, $b_t = (a_c + b_c)/2$, and $c_t = c_c$, where $a_t = a_c/\sqrt{2}$.

[39] S. J. Henderson, O. Shebanova, A. L. Hector, P. F. McMillan, and M. T. Weller, *Structural Variations in Pyrochlore-Structured $Bi_2Hf_2O_7$, $Bi_2Ti_2O_7$ and $Bi_2Hf_{2-x}Ti_xO_7$ Solid Solutions as a Function of Composition and Temperature by Neutron and X-Ray Diffraction and Raman Spectroscopy*, Chem. Mater. **19**, 1712 (2007).

[40] G. Laurita, D. Hickox-Young, S. Husremovic, J. Li, A. W. Sleight, R. MacAluso, J. M. Rondinelli, and M. A. Subramanian, *Covalency-Driven Structural Evolution in the Polar Pyrochlore Series $Cd_2Nb_2O_{7-x}S_x$*, Chem. Mater. **31**, 7626 (2019).

[41] M. Fiebig, V. V Pavlov, and R. V Pisarev, *Second-Harmonic Generation as a Tool for Studying Electronic and Magnetic Structures of Crystals: Review*, J. Opt. Soc. Am. B **22**, 96 (2005).

[42] O. Muller, W. B. White, and R. Roy, *Crystal Chemistry of Some Technetium-Containing Oxides*, J. Inorg. Nucl. Chem. **26**, 2075 (1964).

[43] F. Beech, W. M. Jordan, C. R. A. Catlow, A. Santoro, and B. C. H. Steele, *Neutron Powder Diffraction Structure and Electrical Properties of the Defect Pyrochlores $Pb_{1.5}M_2O_{6.5}$ (M = Nb, Ta)*, J. Solid State Chem. **77**, 322 (1988).

[44] J. Reading, C. S. Knee, and M. T. Weller, *Syntheses, Structures and Properties of Some Osmates(IV,V) Adopting the Pyrochlore and Weberite Structures*, J. Mater. Chem. **12**, 2376 (2002).

[45] U. Dang, J. O'hara, H. A. Evans, D. Olds, J. Chamorro, D. Hickox-Young, G. Laurita, and R. T.

Macaluso, *Vacancy-Driven Disorder and Elevated Dielectric Response in the Pyrochlore $Pb_{1.5}Nb_2O_{6.5}$*, Inorg. Chem. **61**, 18601 (2022).

[46] R. A. Beyerlein, H. S. Horowitz, J. M. Longo, M. E. Leonowicz, J. D. Jorgensen, and F. J. Rotella, *Neutron Diffraction Investigation of Ordered Oxygen Vacancies in the Defect Pyrochlores, $Pb_2Ru_2O_{6.5}$ and $PbTlNb_2O_{6.5}$*, J. Solid State Chem. **51**, 253 (1984).

[47] Y. Hirata, M. Nakajima, Y. Nomura, H. Tajima, Y. Matsushita, K. Asoh, Y. Kiuchi, A. G. Eguiluz, R. Arita, T. Suemoto, and K. Ohgushi, *Mechanism of Enhanced Optical Second-Harmonic Generation in the Conducting Pyrochlore-Type $Pb_2Ir_2O_{7-x}$ Oxide Compound*, Phys. Rev. Lett. **110**, 187402 (2013).

[48] R. D. Shannon, J. D. Bierlein, J. L. Gillson, G. A. Jones, and A. W. Sleight, *Polymorphism in $Bi_2Sn_2O_7$*, J. Phys. Chem. Solids **41**, 117 (1980).

[49] I. Radosavljevic Evans, J. A. K. Howard, and J. S. O. Evans, *α-$Bi_2Sn_2O_7$ - A 176 Atom Crystal Structure from Powder Diffraction Data*, J. Mater. Chem. **13**, 2098 (2003).

[50] W. R. Cook and H. Jaffe, *Ferroelectricity in Oxides of Fluorite Structure*, Phys. Rev. **88**, 1426 (1952).

[51] F. Jona, G. Shirane, and R. Pepinsky, *Dielectric, X-Ray, and Optical Study of Ferroelectric $Cd_2Nb_2O_7$ and Related Compounds*, Phys. Rev. **98**, 903 (1955).

[52] T. C. Kobayashi, Y. Irie, J. Yamaura, Z. Hiroi, and K. Murata, *Superconductivity of Heavy Carriers in the Pressure-Induced Phases of $Cd_2Re_2O_7$*, J. Phys. Soc. Japan **80**, 023715 (2011).

**Supplemental Material**

**Structural analysis for phase II assuming a cubic space group**

For phase I at 350 K, we successfully refined the structure assuming an α-pyrochlore structure in the space group $Fd\bar{3}m$, as shown in Fig. S1(a). In contrast, the crystal structure in phase II at 104 K obtained by structural analysis assumes the cubic space group $F\bar{4}3m$ [Fig. S1(b)] clearly shows the anomalous feature of anisotropic temperature factors in the oxygen sites, which indicates that $F\bar{4}3m$ is not the space group of the Phase II structure. A structural analysis assuming the space group $F23$, the extinction rule of which is the same as that in $F\bar{4}3m$, also failed to improve the anomalous feature in the temperature factor. Since these are the only two space groups in the $Fd\bar{3}m$ subgroup that satisfy the extinction rule, the crystal system in Phase II cannot be cubic.

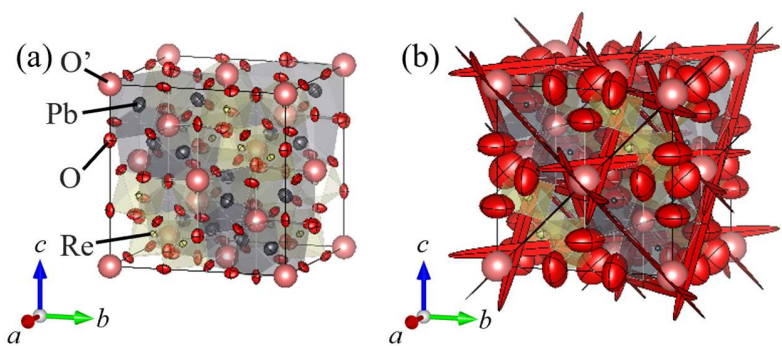

Fig. S1. (a) The crystal structure in phase I at 350 K obtained by structural analysis assuming the space group $Fd\bar{3}m$. Each atom is expressed as the anisotropic temperature factor with a probability of 99%. (d) The crystal structure in phase II at 104 K obtained by structural analysis assumes the cubic space group $F\bar{4}3m$. The anisotropic temperature factor at the O site, shown in red, is apparently incorrect.